\documentclass[prd,
preprint,aps,amsfonts,amssymb,nofootinbib,tightenlines]{revtex4}
\usepackage{amscd}
\usepackage{amsmath}
\usepackage{bm}
\usepackage[dvips]{color}                 %%% for colours in the graphs
\usepackage{graphicx}
\usepackage{eufrak}
\usepackage{eucal}

\input{epsf.sty}

\def\b\mu{{\bf \mu}}

\begin{document}
\preprint{LBNL-53565}

\title{{\Large {\bf Neutral Pion Decay Width in a Hot and Dense Medium}}}

\author{Heron Caldas\footnote{Permanent address: Universidade Federal de Sao Joao del Rey, Departamento de Ci\^{e}ncias Naturais, S\~{a}o Jo\~{a}o del Rei, 36300-000, MG, Brazil; Email address: {\tt hcaldas@ufsj.edu.br}}}
\affiliation{Lawrence-Berkeley Laboratory, Berkeley, CA 94720, U.S.A.}
%\date{August, 2003}

\begin{abstract}
We study the behavior of the $\pi^0$ width as a function of temperature and density. We provide simple expressions for the neutral pion width in a hot and dense medium based on a renormalized version of the microscopic Nambu Jona Lasinio model. Then we consider the two cases where the pion mass is finite or vanishes at the critical temperature and discuss the phenomenological consequences of both cases.
\end{abstract}
\maketitle

\vspace{1cm}

%\pacs{PACS numbers: 11.10.Wx, 12.39.-X, 52.60.+h}

%11.10.Wx Finite-temperature field theory
%12.39.-X Phenomenological quark models
%52.60.+h Relativistic plasma

\vspace{2cm}

\newpage

\section{Introduction}

Particles properties like mass, dispersion relation, lifetime and width, are expected to be modified under extreme conditions such
as high temperature and density. These conditions can be encountered in the nucleus of some stars or fabricated in ultrarelativistic heavy
ion collisions. In this fascinating context one is motivated to deal with a model that is simpler to work with than QCD and that contains
the same symmetry structure of QCD. In this set of models, we choose the NJL model \footnote{Besides renormalizability, another basic reason to deal with the NJL model is that it gives an appropriate understanding of the behavior of the chiral order parameter with temperature which is extracted from lattice simulations of QCD \cite{Karsch}.} \cite{Nambu} in one of its renormalizable extensions \cite{Andre} to investigate the neutral pion decay in a hot and dense medium.

One of the most important contributions to the production of low-energy photons in an ultrarelativistic heavy-ion collision comes from the electromagnetic decay of the neutral pion. To be more precise, photons from $\pi^0$ and $\eta$ decays are responsible for almost $97 \%$ of the inclusive photon spectrum \cite{Aggarwal}. Then, given the importance of such a phenomena, one is motivated to study the modifications experienced by the neutral pions in a hot and dense medium.

The $\pi_0 \to \gamma \gamma$ decay at finite temperature has been studied before by various authors, but we focus here only on the results of Refs. \cite{Hashimoto} and \cite{Bi} which are closely related with ours. Hashimoto {\it et al} \cite{Hashimoto} found that for a finite pion mass at the critical temperature \cite{Hatsuda}, there is an enhancement of the neutral pion decay. On the other hand, Bi Pin-Zhen and Rafelski \cite{Bi} considered a vanishing pion mass at the critical point, which implies in a suppression of neutral pion width near this temperature. We show that, depending only on the solution taken for the vacuum spectation value of the sigma field, we reproduce those results of Refs. \cite{Hashimoto} and \cite{Bi}. We also extend our investigation to finite densities, not considered in Refs. \cite{Hashimoto} and \cite{Bi}.

\section{The renormalized NJL Model and the Minimum of the Classical Potential}

It has been shown that from the canonical form of the NJL model one gets with both the $1/N$ \cite{Shizuya,Weinberg} and mean-field approximations \cite{Andre} an effective renormalizable theory which is like the LSM (linear sigma model) \cite{Gell-Mann}. The reader which is unfamiliar with the LSM see, e.g. Refs. \cite{Koch,Caldas1}. The bosonic part of this effective theory is written as
\begin{equation}
\label{lagsy}
{\cal L}_{sym}  =  \frac{1}{2}\left[(\partial \sigma')^2 + (\partial \vec \pi)^2\right]
 - \frac{\lambda}{4}(\sigma'^2 + \vec\pi^2 - f_\pi^2 )^2  +\cdots \, ,
\end{equation}
where $\sigma'$ and $\pi$ represent the sigma and pion fields, respectively, $\lambda$ is a positive dimensionless coupling constant
and $f_\pi$ is the pion decay constant in vacuum. The Lagrangian density above is symmetric and invariant under an $SU(2)_L \times SU(2)_R$ chiral group.
The vector and axial vector currents associated with this model are given respectively by
\begin{eqnarray}
\vec{V}_{\mu} =  \vec{\pi} \times \partial_{\mu} \vec{\pi},  \\
\nonumber
\vec{A}_{\mu} =  \sigma' \partial_{\mu} \vec{\pi} - \vec{\pi} \partial_{\mu} \sigma' \, ,
\end{eqnarray}
which are conserved. If the up and down quark masses were zero, QCD would have a chiral $SU(2)_L \times SU(2)_R$  symmetry. In the vacuum this symmetry is spontaneously broken with the appearance of the Goldstone bosons. In reality the quark masses are very small but nonzero, so that chiral symmetry is only approximate and the pion has a small mass. An explicit chiral symmetry breaking term is added to the Lagrangian which generates the realistic finite pion mass so that

\begin{equation}
{\cal L' } = {\cal L}_{sym} + {\cal L}_{sym b} \, ,
\label{lagd}
\end{equation}
with

\begin{equation}
{\cal L}_{sym b}  =   c \sigma' \, ,
\label{laga}
\end{equation}
where $c$ is small and positive. The axial vector current is now only ``approximately conserved'', $\partial_{\mu} \vec{A}^{\mu} = c \vec{\pi}$. The minimum of the classical potential in the symmetry broken theory is found to be

\begin{eqnarray}
\label{sol1}
\vec\pi_{0}=0  ,\\
\lambda(\sigma_{0}'^2 - f_\pi^2) \sigma_{0}' = c \, .
\nonumber
\end{eqnarray}

\section{The Neutral Pion Decay Rate and Lifetime in a Hot and Dense Medium}

From the renormalized version of the NJL model \cite{Andre}, one can express the neutral pion width at zero temperature
and density as

\begin{equation}
\Gamma_{\pi^0 \to \gamma\gamma}=\frac{m_{\pi^0}^3}{64 \pi}
F_{\pi \gamma \gamma}^2 \, ,
\end{equation}
where we consider $F_{\pi \gamma \gamma}$ as the formfactor associated with the neutral pion decay into two on-shell photons. In the chiral limit, one arrives at the cutoff independent result

\begin{equation}
\label{width}
\Gamma_{\pi^0 \to \gamma\gamma}= \frac{m_{\pi^0}^3}{64 {\pi}^3} \left( \frac{\alpha_{em}}{f_\pi} \right)^2 =
\frac{m_{\pi^0}^3}{64 \pi} \left( \frac{N_C e^2}{12 \pi^2 f_\pi} \right)^2 \,.
\end{equation}
It is well known that this formula is a result of a relation between the PCAC (partial conservation of axial-vector current) and an anomaly \cite{Jackiw}.
Since it has been proved that the anomaly does not depend on temperature \cite{Itoyama}, the thermal effects should be taken into account through $f_\pi$. We extend this also for finite densities. From Eq. (\ref{sol1}) we get up to first order in $c$,

\begin{equation}
\sigma_{0}' = f_\pi + \frac{c}{2 \lambda f_\pi^2} \equiv \nu \, .
\label{sol2}
\end{equation}
From (\ref{sol2}) we see that $\sigma'$ has a non-zero vacuum expectation value. It is convenient to redefine the sigma field as $\sigma' \to \sigma + \nu$ such that $\sigma$ has zero expectation value. From the shifted Lagrangian the pion mass read

\begin{equation}
m_{\pi}^2=m^2+\lambda\nu^2 \simeq \frac{c}{f_\pi} \, ,
\label{massp}
\end{equation}
where $m^2=-\lambda f_{\pi}^2 <0$. With the experimental value of the pion decay constant $f_\pi$, $c$ has to be adjusted to reproduce the neutral pion mass at zero temperature. In the equation above we have made use of Eq. (\ref{sol2}).

Then we rewrite Eq. (\ref{width}) as
\begin{equation}
\label{width2}
\Gamma_{\pi^0 \to \gamma\gamma}({\rm T},\rho)= \frac{z}{{f_\pi}^{\frac{7}{2}}({\rm T},\rho)}  \,,
\end{equation}
where $z=\frac{ {c}^{\frac{3}{2}} }{64 \pi}\left( \frac{N_C e^2}{12 \pi^2} \right)^2$. In (\ref{width2}) we have used $m_{\pi}$ given by Eq. (\ref{massp}). Thus, in order to know the behavior of the pion properties in extreme conditions such as high temperatures or densities, it is crucial the knowledge of $f_\pi$ as function of these variables \cite{Caldas2}.

\subsubsection{Case I: Finite Pion Mass at the Critical Temperature}

For simplicity, we parametrized $f_\pi({\rm T})$ and $f_\pi(\rho)$ obtained from the NJL model \cite{Bernard}, which are given by

\begin{equation}
\label{fp1}
f_\pi({\rm T})=\frac{a_0}{1+e^{a_1({\rm T}-a_2)}}+a_3  \,,
\end{equation}

\begin{equation}
\label{fp2}
f_\pi(\rho/{\rho_c})=\frac{b_0+b_4~\rho/{\rho_c}}{1+e^{b_1( {\rho/{\rho_c}-b_2) }}}+b_3  \,,
\end{equation}
with $\rho_c=5 \rho_0$, where $\rho_0=0.17 {\rm fm}^{-3}$ is the nuclear matter density. The constants of the expressions above are adjusted such that $f_\pi({\rm T}=0)=f_\pi(\rho/{\rho_c}=0)=92.4 {\rm MeV}$ leading to $m_{\pi^0}({\rm T}=0)=m_{\pi^0}(\rho/{\rho_c}=0)=135 {\rm MeV}$,

\begin{eqnarray}
a_0= 88.9431 ~{\rm MeV} \,, \\
\nonumber
a_1= 0.04167 ~{\rm MeV}^{-1} \,, \\
\nonumber
a_2= 198.6083 ~{\rm MeV} \,, \\
\nonumber
a_3= 3.48 ~{\rm MeV} \,, \\
\nonumber
b_0= 90.4505 ~{\rm MeV} \,, \\
\nonumber
b_1= 3.24  \,, \\
\nonumber
b_2= 6.13  \,, \\
\nonumber
b_3= 1.95  ~{\rm MeV} \,, \\
\nonumber
b_4= -9.38 ~{\rm MeV} \,.
\end{eqnarray}
For the neutral pion lifetime we define
\begin{equation}
\label{lt}
\tau_{\pi^0}({\rm T},\rho)=\frac{1}{\Gamma_{\pi^0 \to \gamma\gamma}({\rm T},\rho)} = \frac{{f_\pi({\rm T},\rho)}^{\frac{7}{2}}}{z} \, ,
\end{equation}
where $\tau_{\pi^0}({\rm T}=\rho/{\rho_c}=0)=8.4 ~\rm{x}~ 10^{-17} \rm{s}$. We note that all meson properties of interest for us in this work are expressed merely as a function of $f_\pi$, in which the properties of the vacuum are encoded. The results for $f_\pi$ and for the pion
mass, width and lifetime as a function of temperature and density for {\it Case I} are shown in the figures (\ref{1}) to (\ref{8}) below. We observe that the pion mass remains almost constant as the chiral transition point is approached and increases after this point, both as a function of temperature and density. This is usually referred as decoupling of the pion from matter \cite{Hatsuda,Bernard}. The neutral pion width is also approximately constant until the phase transition temperature (density), where it grows up, whereas its lifetime tends to zero around this same temperature. These same results have been found in Ref. \cite{Hashimoto} where the triangle diagram \footnote{The triangle diagram at finite temperature has been computed several times in the past years. We will quote only one work which brings a extended list of references and a detailed description of the subtleties associated with the limit of the soft external momenta \cite{Boyanovsky}.} for the $\pi^0 \to \gamma \gamma$ decay has been explicitly calculated.

\begin{figure}[t]
\includegraphics[height=2.2in]{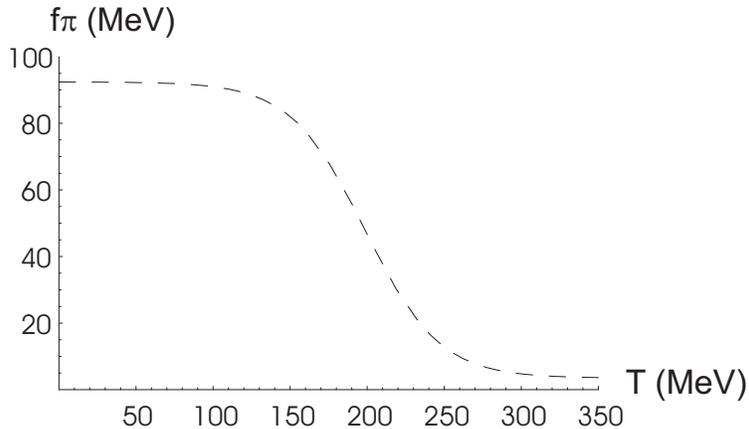}
\caption{\label{1}\textit{  $\rm{f}_{\pi}$ as a function of temperature.  } }
\end{figure}

\begin{figure}[t]
\includegraphics[height=2.2in]{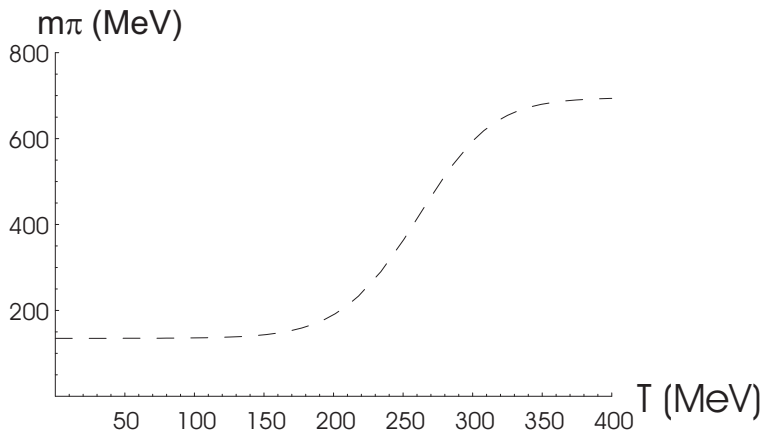}
\caption{\label{2}\textit{  $\rm{m}_{\pi^0}$ as a function of temperature. } }
\end{figure}

\begin{figure}[t]
\includegraphics[height=2.2in]{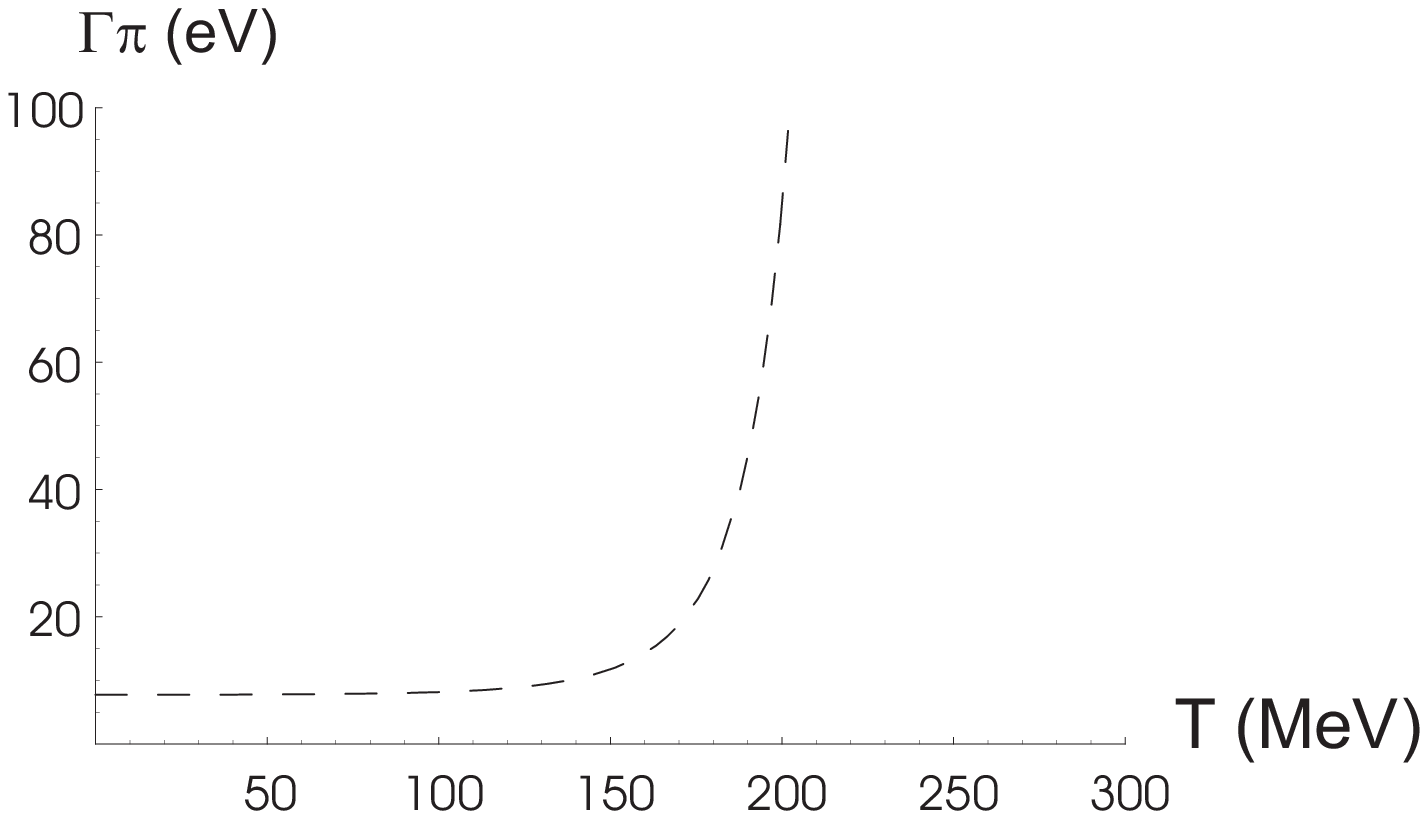}
\caption{\label{3}\textit{  $\pi^0$ width as a function of temperature for Case I.} }
\end{figure}

\begin{figure}[t]
\includegraphics[height=2.2in]{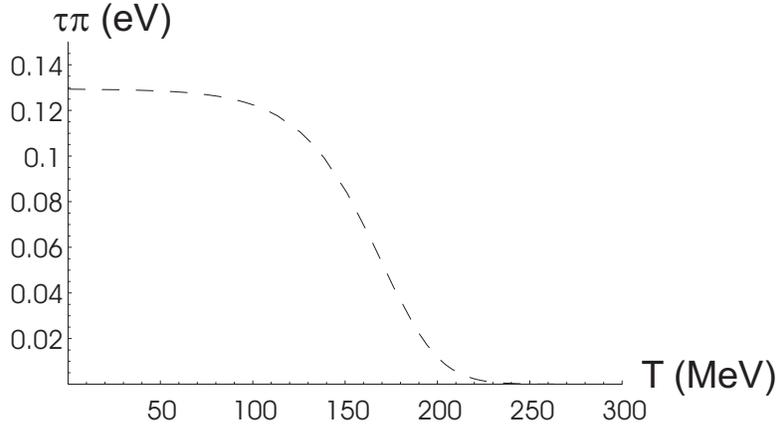}
\caption{\label{4}\textit{  $\pi^0$ lifetime as a function of temperature for Case I.} }
\end{figure}

\begin{figure}[t]
\includegraphics[height=2.2in]{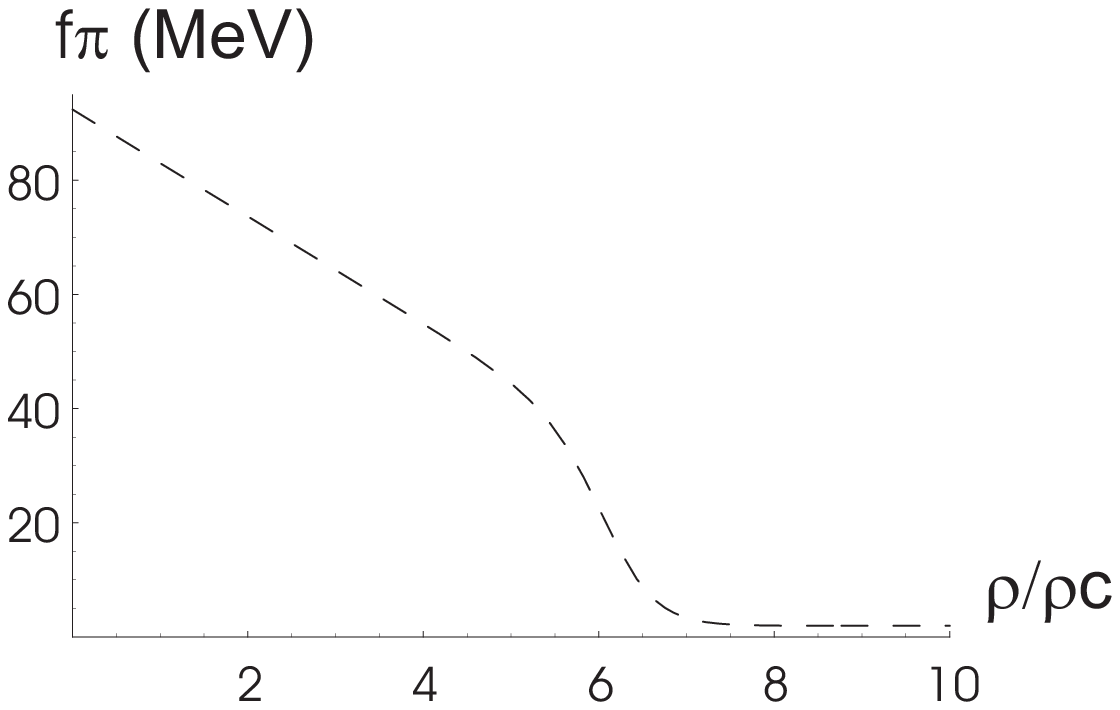}
\caption{\label{5}\textit{  $\rm{f}_\pi$ as a function of the density ratio $\rho / \rho_c$ for Case I.} }
\end{figure}

\begin{figure}[t]
\includegraphics[height=2.2in]{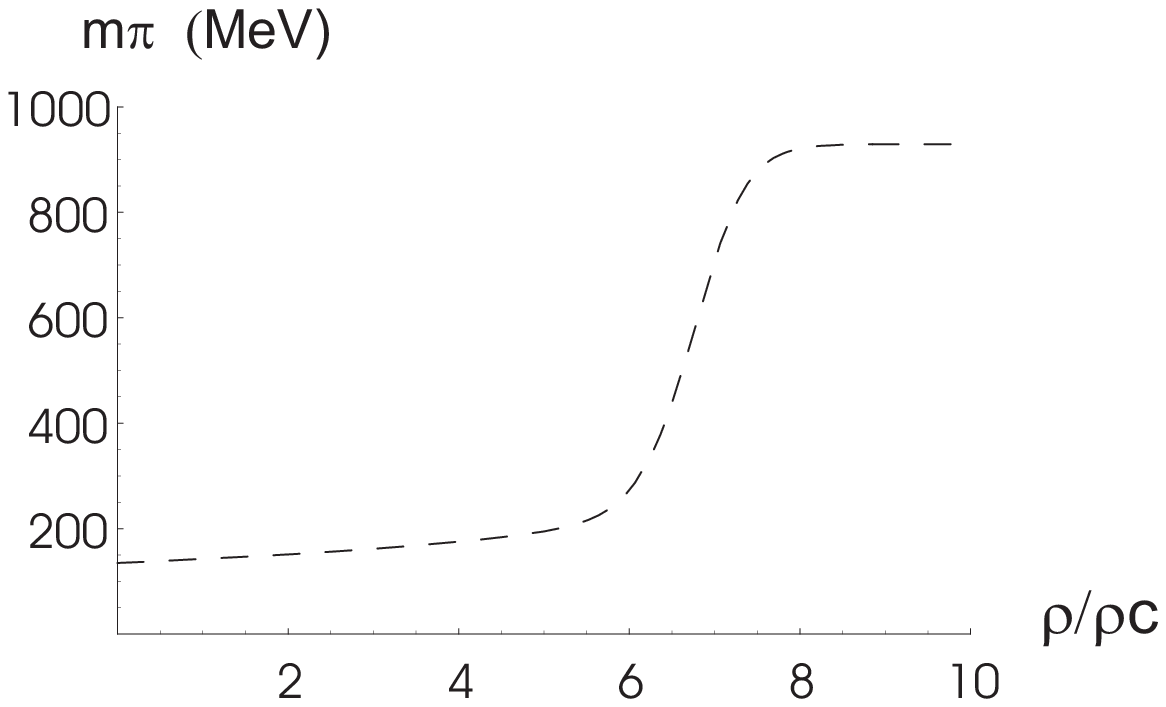}
\caption{\label{6}\textit{  $\rm{m}_\pi^0$ as a function of the density ratio $\rho / \rho_c$ for Case I.} }
\end{figure}

\begin{figure}[t]
\includegraphics[height=2.2in]{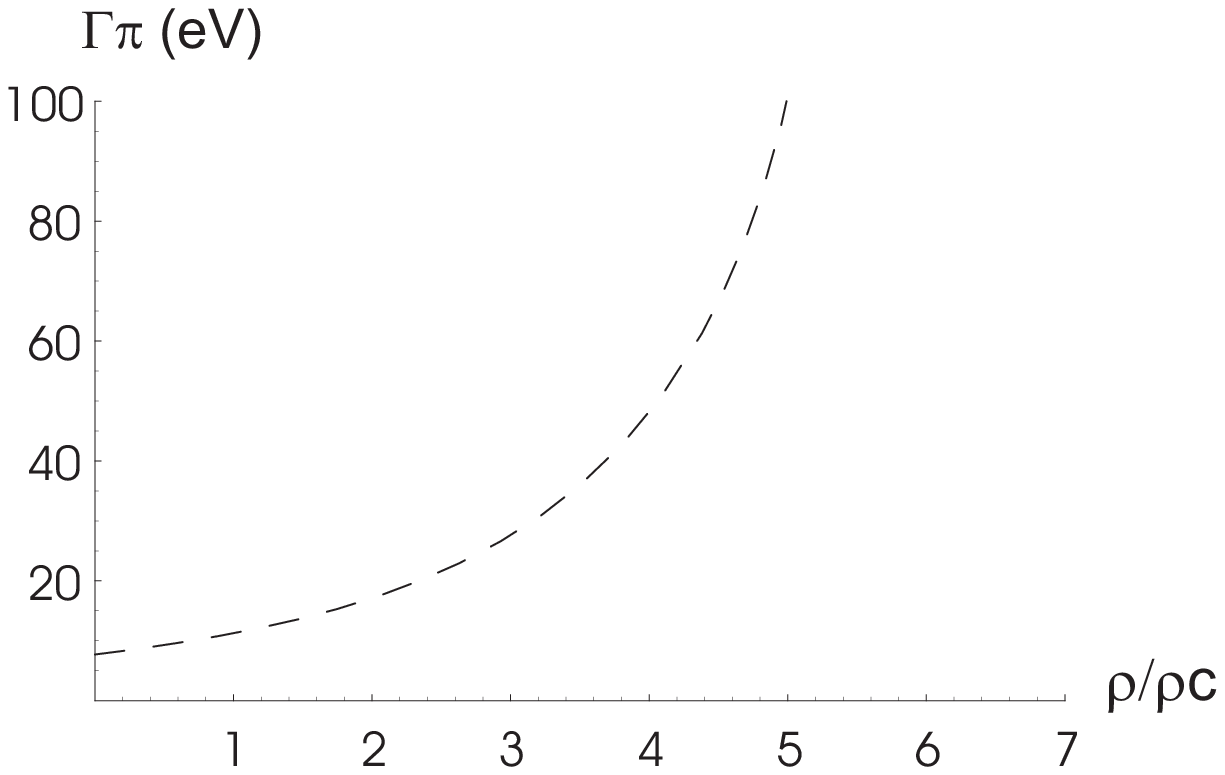}
\caption{\label{7}\textit{  $\pi^0$ width as a function of the density ratio $\rho / \rho_c$ for Case I.} }
\end{figure}

\begin{figure}[t]
\includegraphics[height=2.2in]{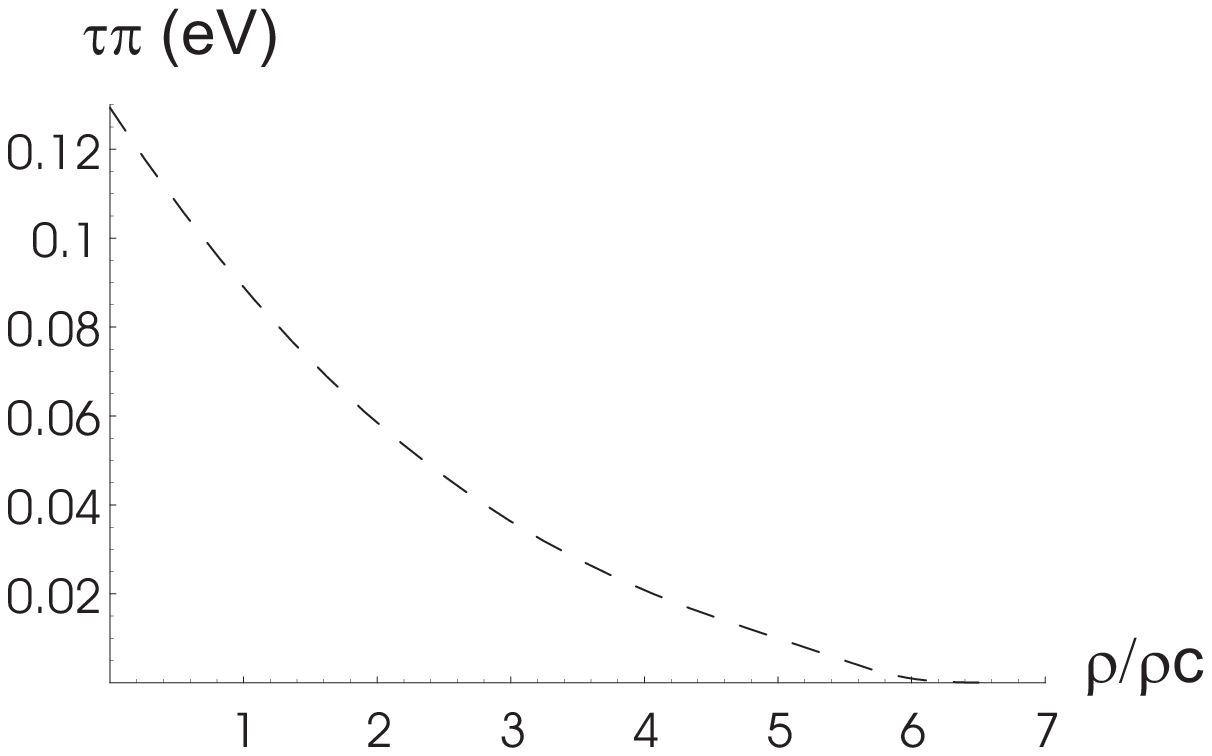}
\caption{\label{8}\textit{  $\pi^0$ lifetime as a function of the density ratio for Case I.} }
\end{figure}

\subsubsection{Case II: Vanishing Pion Mass at the Critical Temperature}

\begin{figure}[t]
\includegraphics[height=2.2in]{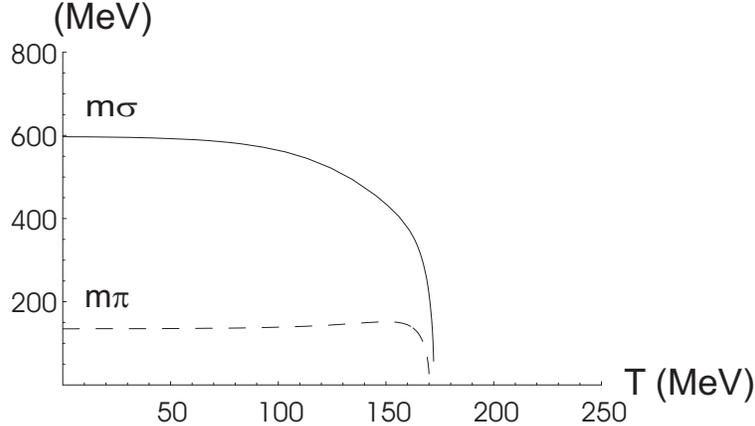}
\caption{\label{9}\textit{ $m_{\sigma}$ and $m_{\pi^0}$ as a function of temperature for case II.} }
\end{figure}

\begin{figure}[t]
\includegraphics[height=2.2in]{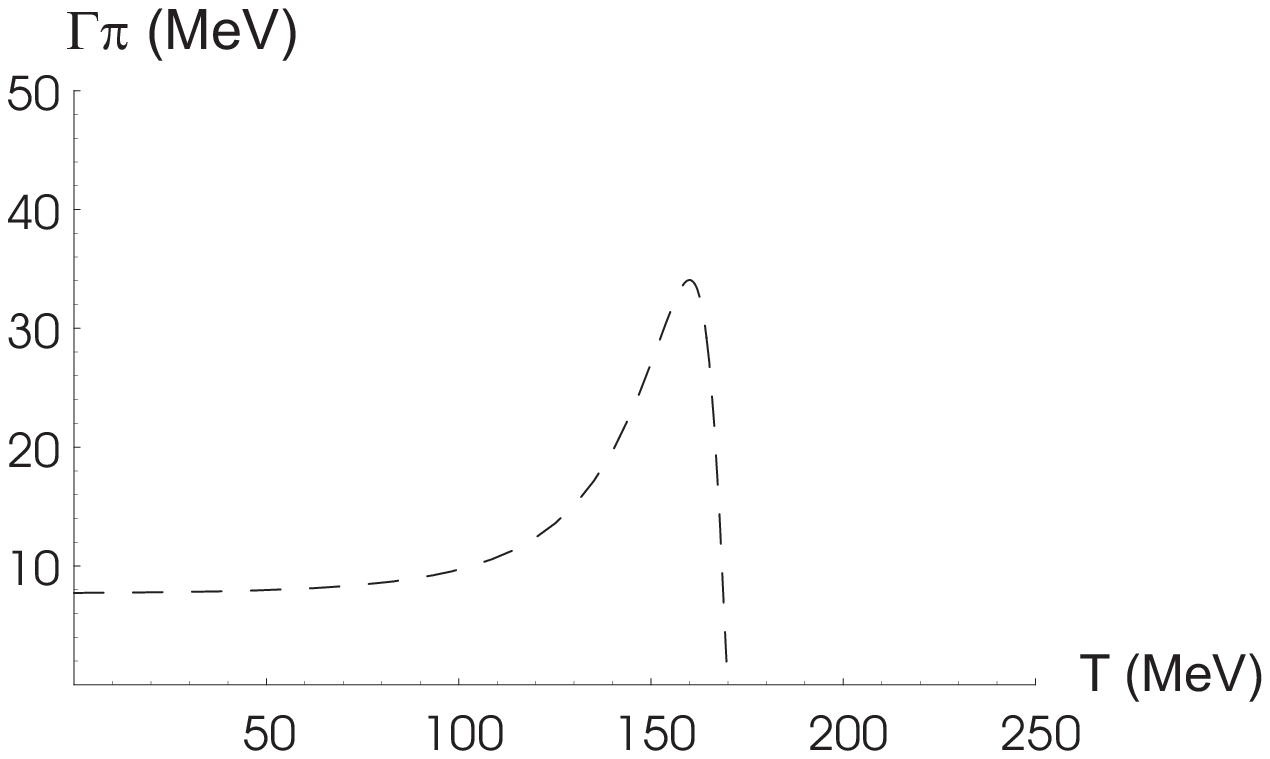}
\caption{\label{10}\textit{ $\pi^0$ width as a function of temperature for case II.} }
\end{figure}

\begin{figure}[t]
\includegraphics[height=2.2in]{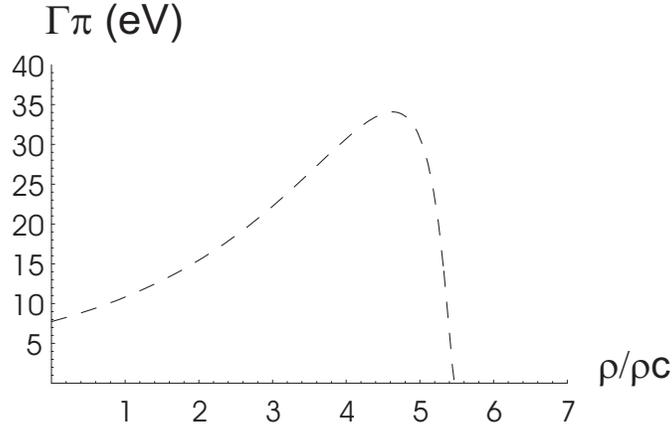}
\caption{\label{11}\textit{ $\pi^0$ width as a function of the density ratio $\rho/\rho_c$ for case II.} }
\end{figure}

Now we turn to the case considered in Ref. \cite{Bi} where the pion mass vanishes at the phase transition temperature. To avoid a divergent $\pi^0$ width near this critical temperature, the authors of Ref. \cite{Bi} allowed for the coefficient of the symmetry breaking term $c$ to be temperature dependent such that, near ${\rm T_c}$, $c({\rm T=T_c})=0$ (where ${\rm T_c}$ is the critical temperature) in order to have a melting of the neutral pion decay width near ${\rm T_c}$. The basic argumentation of the authors of Ref. \cite{Bi} to impose the vanishing of $c(\rm{ T})$ at ${\rm T_c}$ is that if the Gell-Mann-Oakes-Renner \cite{GOR} relation is valid at finite temperature, then it would dictate this behavior for $c(\rm{ T})$. However, $c(\rm{ T})$ as derived in Ref. \cite{Bi} has that features only around ${\rm T_c}$ and under certain conditions. We now discuss how to obtain the same results of Ref. \cite{Bi} without the problems and limitations they found in the derivation of $c({\rm T})$. If we go back to the second equation of (\ref{sol1}), we find up to second order in $c$

\begin{equation}
\label{c1}
\sigma_{0}' \approx f_\pi + \frac{c}{2 \lambda f_\pi^2} -\frac{3 c^2}{8 \lambda^2 f_\pi^5} \equiv \nu (f_\pi)\, .
\end{equation}
Inserting $\nu(f_\pi)$ from this equation in the equations for $m_\pi$ and $m_\sigma$ (keeping up to second order in $c$) extracted form the shifted Lagrangian,

\begin{eqnarray}
m_{\pi}^2({\rm T},\rho)=m^2+\lambda\nu^2(f_\pi) \, , \\
\nonumber
m_{\sigma}^2({\rm T},\rho)=m^2+3\lambda\nu^2 (f_\pi)\, ,
\label{masses}
\end{eqnarray}
we find the values of $c$ and $\lambda$ that give the experimental value of the pion and sigma masses at zero temperature and density. We have used $m_\sigma({\rm T} = \rho/\rho_c =0)= 600 {\rm MeV}$. For this case the neutral pion width in the medium is
\begin{equation}
\label{width3}
\Gamma_{\pi^0 \to \gamma\gamma}({\rm T},\rho)= \frac{1}{64 \pi} \left( \frac{N_C e^2}{12 \pi^2 } \right)^2 \frac{m_{\pi^0}^3({\rm T},\rho)}{f_\pi^2({\rm T},\rho)} \, ,
\end{equation}
with $f_\pi({\rm T},\rho)$ as before given by Eqs. (\ref{fp1}) and (\ref{fp2}) respectively. The constants necessary to reproduce the vanishing of $f_\pi({\rm T},\rho)$ near the critical point (again parametrized from Ref. \cite{Bernard}) read as
\begin{eqnarray}
a_0= 132.57 ~{\rm MeV} \,, \\
\nonumber
a_1= 0.037503 ~{\rm MeV}^{-1} \,, \\
\nonumber
a_2= 178.747 ~{\rm MeV} \,, \\
\nonumber
a_3= -40 ~{\rm MeV} \,, \\
\nonumber
b_0= 92.4042 ~{\rm MeV} \,, \\
\nonumber
b_1= 3.24  \,, \\
\nonumber
b_2= 5.517  \,, \\
\nonumber
b_3= 0  ~{\rm MeV} \,, \\
\nonumber
b_4= -9.38 ~{\rm MeV} \,.
\end{eqnarray}
The behavior of the meson masses as well as the neutral pion width  as a function of temperature and density for {\it Case II} can be seen in figures (\ref{9}) to (\ref{11}).

\section{Conclusions}
\label{conc}

We have studied the behavior of some static properties (mass, width and lifetime) of the neutral pion under extreme conditions like high temperature
and density in the framework of a renormalized version (extension) of the NJL model which is basically the LSM. We provided simple expressions for this analysis based on the fact that $f_\pi$ take into account the modifications imposed by the medium.

We have found that depending on the solution chosen for the vacuum expectation value of the sigma field (i.e., depending on the investigated scenario), we reproduce those results of Refs. \cite{Hashimoto} (finite pion mass at the critical point) and \cite{Bi} (vanishing pion mass at the critical point). As the true vacuum is only one, the final word on this subject belongs to the experiment. Thus, in order to get a conclusive answer for this question, it is crucial to observe and analyze the hadron and photon spectra near the critical temperature in a heavy-ion collision experiment.

We note that if {\it Case II} (vanishing pion mass at the critical point) is correct, then there would be no alteration in the $\pi^0 \to \gamma \gamma$ decay rate in hot or dense matter near ${\rm T_c}~({\rho_{\rm c}})$. This is because the hot hadronic matter lifetime is $\approx 10 fm/c$ which is much less than the (rest) neutral pions lifetime that are $\approx 10^7~ fm/c$. In this case the neutral pions would mostly decay in the final state (i.e., after freeze-out).

In the other scenario ({\it Case I}), where the neutral pion decay is enhanced near the critical temperature, some of the pions that are located in regions of the hadronic matter near to ${\rm T_c}~({\rho_{\rm c}})$ would decay inside the plasma and their width would be affected by it. Such an increase might be detected in heavy-ion collisions.

It is worth to point out here that the present photon momentum resolution of the STAR experiment~\cite{data1} does not allow any decisive conclusions about the possible enhancement of the $\pi^0$ width\footnote{See also Refs. \cite{data2,data3} for the $\pi^0$ spectra via the $\pi^0 \to \gamma \gamma$ decay by the PHENIX experiment.}.

Then, is seems that the answer to the question of whether there is an enhancement or suppression of the $\pi^0 \to \gamma \gamma$ decay rate at ${\rm T_c}~({\rho_{\rm c}})$ still depend on experimental evidence.

A natural extension of this work would be the verification of the ideas presented here in other processes where the decay rates also depend only on $f_{\pi}$ and the masses involved (such as in ~\cite{Kapusta}), with these quantities expressed as a function of temperature and density derived accordingly within the model under consideration.

\begin{acknowledgments}

The author would like to thank A.L. Mota, A. Majumder, X.N. Wang, J. Kapusta, D. Boyanovsky and S. Soff for useful discussions and P. Bedaque for helpful discussions and for a critical reading of the manuscript. I also thank S. Klein for illuminating conversations about the STAR experiment. The author thanks the Nuclear Science Division of LBL for hospitality and support. This work was partially supported by CAPES/Brazil.

\end{acknowledgments}

%% \bibliographystyle{h-physrev4}

%\begin{references}

%\end{references}


\begin{thebibliography}{10}


\bibitem{Nambu}  Y. Nambu and G. Jona-Lasinio, {Phys. Rev. }{\bf 122}, 345 (1961).

\bibitem{Karsch} F. Karsh, Quark-gluon plasma, R.C. Hwa, ed. (World Scientific, Singapore, 1993) p. 61.

\bibitem{Andre}  A.L. Mota, M.C. Nemes, B. Hiller and H. Walliser, Nucl. Phys. {\bf A652}  73 (1999), hep-ph/9901455.
%%CITATION = HEP-PH 9901455;%%

\bibitem{Aggarwal} M.M. Aggarwal {\it et al.}, Phys. Rev. Lett. {\bf 85}, 3595 (2000).

\bibitem{Hashimoto} T. Hashimoto, K. Hirose, T. Kanki and O. Miyamura Phys. Rev. {\bf D37}, 3331 (1988).
%%CITATION = PHRVA,D37,3331;%%

\bibitem{Bi} Bi Pin-Zhen and J. Rafelski, Mod. Phys. Lett. {\bf A7}, 2493 (1992).

\bibitem{Hatsuda} T. Hatsuda and T. Kunihiro, Phys. Lett. {\bf B185}, 304 (1987); S. Gottlieb, W. Liu, D. Toussaint, R.L. Renken and R.L. Sugar, Phys. Rev. Lett. {\bf 59}, 1881 (1987).

\bibitem{Shizuya} K. Shizuya, Phys. Rev. {\bf D21} 2327 (1980) and references therein.
%%CITATION = PHRVA,D21,2327;%%

\bibitem{Weinberg} S. Weinberg, Phys. Rev. {\bf D56} 2303 (1997).
%%CITATION = PHRVA,D56,2303;%%

\bibitem{Gell-Mann}  M. Gell-Mann and M. Levy, Nuovo Cimento {\bf 16}, 705 (1960).

\bibitem{Koch} V. Koch,  Int. J. Mod.Phys. {\bf E6}, 203 (1997), nucl-th/9706075.
%%CITATION = NUC-TH 9706075;%%

\bibitem{Caldas1} H.C.G. Caldas, A.L. Mota and M.C. Nemes, Phys. Rev. {\bf D63}, 056011 (2001), hep-ph/0005180.
%%CITATION = HEP-PH 0005180;%%

\bibitem{Jackiw} R. Jackiw, {\it Lectures on Current Algebra and Its Application} (Princeton University Press, Princeton, NJ, 1972).

\bibitem{Itoyama} H. Ytoyama and A.H. Mueller, Nucl. Phys. {\bf B218} 349 (1983).

\bibitem{Caldas2} H.C.G. Caldas, D.H.T. Franco, A.L. Mota, F.A. Oliveira and M.C. Nemes, Nucl. Phys. {\bf A617}  464 (1997).

\bibitem{Bernard} V. Bernard, Ulf-G. Meissner and I. Zahed, Phys. Rev. {\bf D36}, 819 (1987).
%%CITATION = PHRVA,D36,819;%%

\bibitem{Boyanovsky} S.P. Kumar, D. Boyanovsky, H.J. de Vega and R. Holman, Phys. Rev. {\bf D61} 065002 (2000), hep-ph/9905374.
%%CITATION = HEP-PH 9905374;%%

\bibitem{GOR} M. Gell-Mann, R.J. Oaks and B. Renner, Phys. Rev. {\bf 175}, 2195 (1968); L.J. Reinders, H. Rubinstein and S. Yazaki, Phys. Rep. {\bf 127}, 1 (1985); J. Gasser and H. Leutwyler, Phys. Rep. {\bf 87}, 77 (1982).

\bibitem{data1} S. Klein (private communication).

\bibitem{data2} K. Adcox {\it et al.}, Phys. Rev. Lett. {\bf 88}, 022301 (2002), nucl-ex/0109003.
%%CITATION = NUC-EX 0109003;%%

\bibitem{data3} K. Adcox {\it et al.}, Phys. Rev. Lett. {\bf 88}, 242301 (2002), nucl-ex/0112006.
%%CITATION = NUC-EX 0112006;%%

\bibitem{Kapusta} J. Kapusta and I. Shovkovy,\newblock Phys. Rev. {\bf C68} 014901 (2003), nucl-th/0209075.
%%CITATION = NUC-TH 0209075;%%

\end{thebibliography}
\end{document}